\documentclass[runningheads]{llncs}

\usepackage[T1]{fontenc}
\usepackage{graphicx}
\usepackage{hyperref}
\usepackage{color}
\usepackage[colorinlistoftodos]{todonotes}
\usepackage{amsmath}
\usepackage{booktabs}

\begin{document}

\title{IoT and Older Adults: Towards Multimodal \\EMG and AI-Based Interaction with Smart Home}

\titlerunning{IoT and EMG: Multimodal Interaction with SmartHome}

\author{Wiesław Kopeć\inst{1,5}\orcidID{0000-0001-9132-4171} \and 
Jarosław Kowalski \inst{2}\orcidID{0000-0002-1127-2832}\and
Aleksander Majda \inst{3}\orcidID{0009-0001-2292-5181}\and
Anna Duszyk-Bogorodzka\inst{4}\orcidID{0000-0002-5984-2339} \and
Anna Jaskulska \inst{5}\orcidID{0000-0002-2539-3934} \and
Cezary Biele \inst{2}\orcidID{0000-0002-2539-3934}
}

\authorrunning{W. Kopec et al.}
\institute{XR Center, Polish-Japanese Academy of Information Technology
\email{kopec@pja.edu.pl}\\
\and Laboratory of Interactive Technology, National Information Processing Institute
\and
Big Idea Technology \and
Institute of Psychology, SWPS University \and
Kobo Association
}

\maketitle              
\begin{abstract}
We report preliminary insights from an exploratory study on non-standard non-invasive interfaces for Smart Home Technologies (SHT). 
This study is part of a broader research project on effective Smart Home ecosystem Sagacity that will target older adults, impaired persons, and other groups disadvantaged in the main technology discourse. Therefore, this research is in line with a long-term research framework of the HASE research group (Human Aspects in Science and Engineering) by the Living Lab Kobo.  
In our study, based on the prototype of the comprehensive SHT management system Sagacity, we investigated the potential of bioelectric signals, in particular EMG and EOG as a complementary interface for SHT. Based on our previous participatory research and studies on multimodal interfaces, including VUI and BCI, we prepared an in-depth interactive hands-on experience workshops with direct involvement of various groups of potential end users, including older adults and impaired persons (total 18 subjects) to explore and investigate the potential of solutions based on this type of non-standard interfaces. The preliminary insights from the study unveil the potential of EMG/EOG interfaces in multimodal SHT management, alongside limitations and challenges stemming from the current state of technology and recommendations for designing multimodal interaction paradigms pinpointing areas of interest to pursue in further studies.


\end{abstract}

\section{Research context}
One of the main global challenges is a demographic transition related to an aging population, longer lives, and smaller families \cite{rudnicka2020world}. One of the promising approaches towards counterparting that trend is to support longer and independent living of older adults with the use of new technologies, including Smart Home, Internet of Things (IoT) and automation. This approach underpins a concept of Ambient and Assisted Living \cite{sun2009promises} or Active and Assisted Living (AAL)\footnote{https://www.aal-europe.eu/}.

However, these new sustainable and pervasive technologies require effective interfaces for management and control towards effective interaction with new technologies by a diverse group of users, including older adults and impaired persons. In this work, we present preliminary results of an exploratory study of interaction with Smart Home Technology (SHT) using an innovative facial expression recognition based on electromyograph (EMG) activity. This research is part of a comprehensive research and development project on providing effective methods and tools towards multimodal interfaces for a universal and open SHT ecosystem. This research is also in line with our long-term studies on multimodal interfaces and interaction, including non-standard and non-invasive interfaces as a part of a research framework HASE group by the Living Lab Kobo, based on participatory research on human-technology interaction, in particular led by Laboratory of Interactive Technologies, National Information Processing Institute (LIT NIPI), XR Center Polish-Japanese Academy of Information Technologies (XRC PJAIT) and Institute of Psychology, SWPS University.

\textbf{Aging population and demographic transition.} One of the main contemporary social and economical challenges is related to the global demographic transition, based on growing trends of aging the population, longer lives and smaller families \cite{united2024world,rudnicka2020world}. This is valid especially in developed countries, including EU, North America, and Japan \cite{eurostat2020europe}. However, according to numerous reports at the international and national level, including United Nations and European Union agencies, this is an irreversible global trend that affects even countries with relatively youthful populations.\cite{united2024world} In 2021 one in ten people worldwide were aged 65 or above, with the prospect of 1 in 6 globally by the half of this century. 
The global population aged 65 and older is projected to more than double by the 2050, reaching 1.6 billion, while the number of people aged over 80 years old is growing even faster \cite{united2023world} .
In fifty years, i.e. by the late 2070s, the number of older adults over 65 worldwide is projected to reach 2.2 billion, exceeding the number of the population under 18. However, by the mid-2030s, older adults aged 80 and over will outnumber infants (1 year of age or less), reaching 265 million \cite{united2024world}. 

\textbf{The Internet of Things and Smart Home Technology} are one of the strongest trends in the modern technology market.
Smart home refers to a residence equipped with smart technologies that provide tailored services for users \cite{marikyan2019SHTsystematicUSER}
based on Internet of Things (IoT) \cite{pal2018IoT_SH_TAM} devices and solutions.
According to recent reports and market analysis from leading market analytics the total IoT market in 2023 was over US \$800 billion and will reach in 2024 almost one trillion USD with a prospect of further growth in the coming years \cite{statista2024iot}. Constant growth in intelligent or smart devices, as well as network technologies, including new standards of ubiquitous protocol 5G, paved the way to the expansion of IoT in various areas, including Industrial IoT, automotive, smart cities, and homes. The latter area, smart home, is a subject of a particularly intensive growth of application, based predominantly on well established residential networking standards (local area networks and broadband Internet connection) and the pervasive consumer IoT market, which has over half of estimated total number of connected IoT devices, namely over 10 billion out of a total of over 18 billion connections in 2024 estimation, with the prospect of exceeding 20 billion out of almost 40 billion in 2028 \cite{statista2024iot}.

\textbf{Participatory Design and Living Lab Approach}
These above mentioned social and technological trends paved the way for the active development of sustainable solutions that will help people stay independent and active for longer in their older age. These endeavors are covered by the umbrella term Ambient and Assisted Living or Active and Assisted Living (AAL) \cite{sun2009promises,jovanovic2022AALumbrella}. AAL postulates making use of recent advancements in technology, including the Internet of Things, smart environments, and automation. Such technological advances can also pose challenges to users, which is why an effective user-centric approach is necessary to develop solutions that are better suited to different end-user groups \cite{pal2018IoT_SH_TAM}. One of the well established approaches is participatory design, or co-design, with direct involvement of end users in the process of co-creation of new solutions better tailored to their needs. This approach is a crucial aspect of human-computer or human-technology interaction research, conducted also by our transdisciplinary HASE research group \cite{kopec2021pdl}. 
A living lab approach is an example of well established experimental approach deeply rooted in the philosophy of user-centered research that fosters the process of designing new, useful, and acceptable products and services. The original Living Lab user-centric concept was coined at MIT MediaLab by William Mitchell \cite{eriksson2005livinglab} over two decades ago, at the advent of IoT, Smart Home and other emerging technologies related to the research reported in this paper. This concept has been effectively explored by our Human Aspects in Science and Engineering research group by the Living Lab Kobo Association, which enables transdisciplinary collaboration between scientists, practitioners, entrepreneurs and volunteers \cite{kopec2021pdl}.

\textbf{Multimodal Interfaces and Interaction}
Having in mind that Smart Home Technology is a well established and constantly growing market, there is a challenge of providing effective interaction towards end users' interaction with new technologies \cite{pal2018IoT_SH_TAM}. Well established concepts of human-computer interfaces like WIMP (Windows, Icons, Mouse, Pointer) \cite{van1997postWIMP} have limited application to the context of smart home environments. 
In the study reported in this paper, we have built on our previous extensive research of our HASE Research group on new interfaces, modalities, and paradigms of interaction in the context of SHT, from Voice Assistants (VA) and Voice user interfaces (VUI) for older adults \cite{kowalski2019vui}, impaired persons \cite{jaskulska2020loss}, and children \cite{biele2019children} to Brain Computer Interfaces (BCI) \cite{kopec2021bci} in the context of experimental participatory research approach by the distributed Living Lab Kobo \cite{kopec2021pdl,kopec2018spiral,karpowicz2021framework,skorupska2021engagement}, 
Therefore, in the course of our research endeavors we examined the potential of non standard interfaces and modes of user-technology interaction based on psychophysiological and bioelectric signals. In particular, the study reported in this paper is a part of our research and development project on the Sagacity Smart Home solution, where we examined various devices and approaches towards effective EMG/EOG interface, including ML/AI algorithms. 

\section{Exploratory Interaction Study}
Based on our previous research and above mentioned research context, as well as state-of-the-art solutions, we have conducted an exploratory experimental session with a working prototype of the comprehensive SHT management system Sagacity coupled with an EMG interface for bioelectric signals based on facial muscles activity and specific facial expressions for intentional communication with the SHT system Sagacity. Following the Living Lab approach, we set up an in-depth exploratory session with end users direct involvement. This session provided hands-on experience to gather observations and insights from the end users group. 

\begin{figure}
    \centering
    \includegraphics[width=1\linewidth]{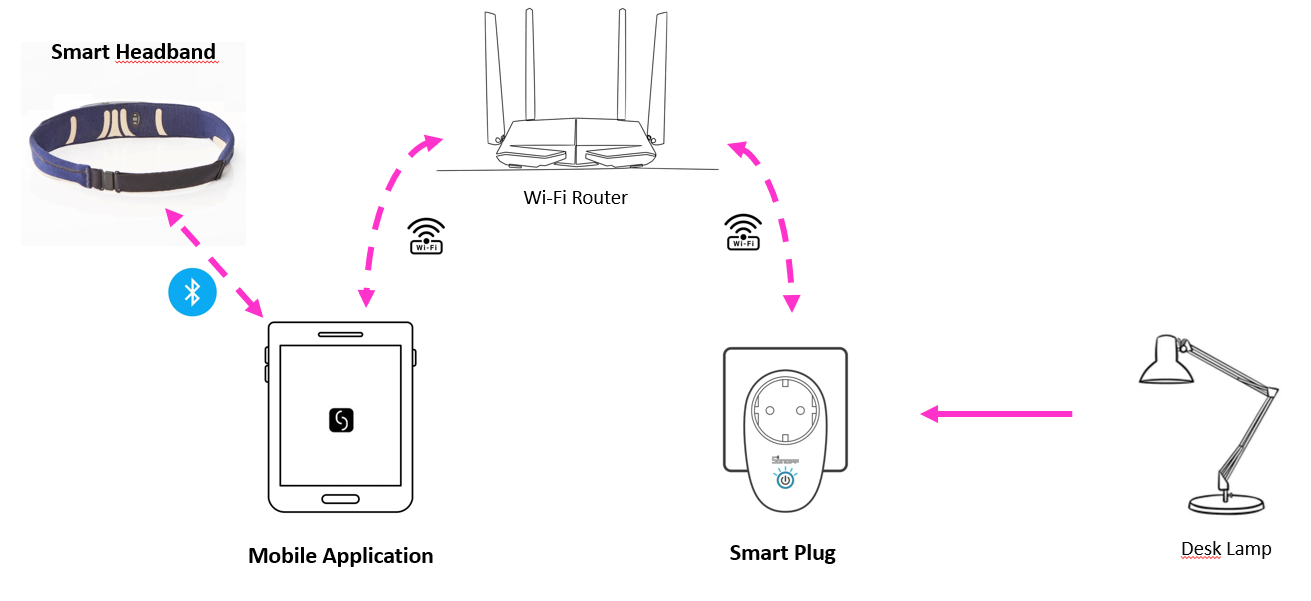}
    \caption{Experimental setup and connections (source: own elaboration)}
    \label{fig:flowDiagram}
\end{figure}

We prepared an experimental setup that provides ecologically valid insight of the various target groups, including older adults and impaired persons. The experimental setup was based on the real-life configuration (fig. \ref{fig:flowDiagram}) that represents the basic SHT configuration. The experimental setup consisted of all necessary elements of a working prototype in the smart home environment of the system, including the smart band paired with a mobile device (tablet) equipped with the Sagacity application that controlled the plain light source via Smart Plug using standard Wi-Fi protocols (table \ref{tab:listofEquipment}). 

\begin{table}[!ht]
    \centering
    \caption{List of equipment in experimental setup}
\label{tab:listofEquipment}

    \begin{tabular}{lll}
        \textbf{Type} & & \textbf{Specs} \\ \hline
        Smart Headband & & Muse S Headband wireless with BT \\ \hline
        Smart Plug & & Sonoff S26 Wi-Fi Smart Plug \\ \hline
        Light source & & traditional desk lamp (Maxcom ML110) \\ \hline
        Mobile Device & & Samsung Galaxy Tab S7 FE with Sagacity application \\ \hline
        WiFi Router & & Cudy WR1200 (1200Mb/s a/b/g/n/ac) DualBand \\ \hline
    \end{tabular}
\end{table}

The prototype of the Sagacity system used in this study is based on three classification algorithms that were verified, trained, and implemented in the previous stages of the Sagacity project and verified towards effective facial expression recognition (Linear Discriminant Analysis (LDA), K-Nearest Neighbors (KNN), and Logistic Regression). These algorithms were used to categorize signal segments into two classes: movement and rest. Feature selection was guided by both existing research \cite{zhang2013emg} and the transdisciplinary research team's expertise. In this paper, we focus on hands-on user experience exploration and the insights gained, as a part of a broader research that will be reported in a separate paper.

The experimental procedure was based on the participatory living lab approach and consisted of hands-on experience sessions with in-depth observations and interviews with the end-users from the target groups. The experimental study involved 18 persons, including six persons between 45 and 65 years old (three male and female participants), six persons over 65 years old (two male and four female participants) and six persons with diagnosed mobility disability, including persons using wheelchairs or crutches (three female and three male participants). The experimental protocol consisted of two major phases: setting up the Sagacity SHT ecosystem and an exploratory hands-on experience session of the SHT Sagacity environment using an EMG interface. In this study, we focus on EMG and AI-based interaction with the pretrained Sagacity system providing preliminary insights from the users and observations that form basis for further exploration, experimental studies and system development.

\section{Preliminary Results and Insights}

In general, all users were able to operate and control the preconfigured experimental EMG SHT setup, specifically controlling the light using the preliminary working prototype of Sagacity.
Insights and observations from this preliminary experimental study of the working prototype of comprehensive EOG/EMG SHT system Sagacity are in line with information gathered at the previous steps of the participatory design process with users, i.e. older adults from our Living Lab Kobo, which are also in line with the project assumptions. In particular, users pointed out that the potential target groups of the EMG-controlled SHT system may be older adults and impaired persons (including temporary and permanent impairment). On the other hand, users also pointed out that based on the ease of use of the smartband, the system may be targeted to various end users groups, as a part of the multimodal system, that will support users as a primary or supplementary interface according to the circumstances. In particular, users identified EMG-based control as the primary or preferred interface, especially in situations where they need hands free or due to privacy concerns (when multiple persons share the same space). They also pointed out that such EMG interfaces may be treated as supplementary in typical scenarios like controlling the light or calling the assistance, especially by the impaired persons. These insights were confirmed in the experimental prototype study group. For example, one of the participants, who is responsible for taking care of the permanently disabled older member of the family (paralyzed from the neck down), finds the proposed solution as very promising and useful in everyday caregiving practice. However, this experimental study also unveiled some limitations and challenges related to interface accuracy, the onboarding process (user training time), and the intuitiveness of the interaction paradigms. These areas will be the subject of further research and publications.

\section{Conclusions and Further Work}

The presented exploratory study, as part of the broader research, provides preliminary findings and valuable insights for further research on innovative, non-invasive interfaces for Smart Home Technologies (SHT). By leveraging bioelectric signals, specifically electromyography (EMG) and electrooculography (EOG), as complementary interfaces, we aim to expand the accessibility and usability of SHT. In particular, this research, as part of the Sagacity research and development project, aims to empower older adults, impaired persons, and other marginalized groups within the technology landscape. 
Through a participatory living lab approach involving diverse end-users, we investigated the potential and limitations of these non-standard interfaces. Our findings reveal promising opportunities for multimodal SHT management that are in line with our previous research on human-technology interaction, while highlighting challenges related to the current technological limitations and user acceptance. Future research directions include refining interface accuracy, reducing user training time, and developing intuitive interaction paradigms to fully realize the potential of bioelectric interfaces in SHT. 
By addressing these challenges and capitalizing on the strengths of EMG/EOG interfaces, we can create more inclusive, versatile and user-centric Smart Home environments, which is one of the key aspects of Sagacity project and HASE research group objectives.

% \begin{credits}
% \subsubsection{\ackname}
\section*{Acknowledgments}
The study was conducted as a part of a research and development grant: Development of an innovative system for controlling Internet of Things devices using a wireless bioelectric signal amplifier (grant no. Rzeczy są dla ludzi/0079/2020) by The National Center for Research and Development. However, this study constitutes an example of a bottom-up participatory research initiative done in the spirit of transdisciplinary living lab collaboration between scientists, practitioners, entrepreneurs, and volunteers. Therefore, we would like to thank the many people and institutions gathered together by the Living Lab Kobo and the HASE Research Group. First, we would like to thank all the members of the HASE research group (Human Aspects in Science and Engineering) and the Living Lab Kobo community for their support of this research, especially the older adults, for supporting recruitment and participation in the lab studies.
% \subsubsection{\discintname}
The authors have no competing interests to declare that are relevant to the content of this article.
% \end{credits}

\bibliographystyle{bibliography/splncs04}
\bibliography{bibliography/bibliography}

\end{document}